**Identifying latent shared mobility preference segments in low-income communities: ride-hailing, fixed-route bus, and mobility-on-demand transit**

**Xinyi Wang**
School of Civil and Environmental Engineering
Georgia Institute of Technology
790 Atlantic Drive, Atlanta, GA 30332
Email: xinyi.wang@gatech.edu
ORCID: https://orcid.org/0000-0002-3564-9147

**Xiang Yan, Ph.D.**
Department of Civil and Coastal Engineering
University of Florida
1949 Stadium Rd, Gainesville, FL 32611
Email: xiangyan@ufl.edu
ORCID: https://orcid.org/0000-0002-8619-0065

**Xilei Zhao, Ph.D.**
Department of Civil and Coastal Engineering
University of Florida
1949 Stadium Rd, Gainesville, FL 32611
Email: xilei.zhao@essie.ufl.edu
ORCID: https://orcid.org/0000-0002-7903-4806

**Zhuoxuan Cao**
Department of Civil and Coastal Engineering
University of Florida
1949 Stadium Rd, Gainesville, FL 32611
Email: zhuoxuancao@ufl.edu
ORCID: https://orcid.org/0000-0002-3247-1291

**ABSTRACT**

Concepts of Mobility-on-Demand (MOD) and Mobility as a Service (MaaS), which feature the integration of various shared-use mobility options, have gained widespread popularity in recent years. While these concepts promise great benefits to travelers, their heavy reliance on technology raises equity concerns as socially disadvantaged population groups can be left out in an era of on-demand mobility. This paper investigates the potential uptake of MOD transit services (integrated fixed-route and on-demand services) among travelers living in low-income communities. Specially, we analyze people's latent attitude towards three shared-use mobility services, including ride-hailing services, fixed-route transit, and MOD transit. We conduct a latent class cluster analysis of 825 survey respondents sampled from low-income neighborhoods in Detroit and Ypsilanti, Michigan. We identified three latent segments: **shared-mode enthusiast**, **shared-mode opponent**, and **fixed-route transit loyalist**. People from the **shared-mode enthusiast** segment often use ride-hailing services and live in areas with poor transit access, and they are likely to be the early adopters of MOD transit services. The **shared-mode opponent** segment mainly includes vehicle owners who lack interests in shared mobility options. The **fixed-route transit loyalist** segment includes a considerable share of low-income individuals who face technological barriers to use the MOD transit. We also find that males, college graduates, car owners, people with a mobile data plan, and people living in poor-transit-access areas have a higher level of preferences for MOD transit services. We conclude with policy recommendations for developing more accessible and equitable MOD transit services.

**Keywords**: mobility on demand (MOD), Mobility as a Service (MaaS), transport equity, latent class cluster analysis, ride-hailing, public transit

# 1. INTRODUCTION

New and shared mobility options such as carsharing, bikesharing, and ride-hailing have quickly proliferated across cities. The rapid rise of these privately-operated shared mobility services raises the question of how they interact with public transportation. While some worry that emerging mobility services will disrupt the operation of publicly funded transportation services, a growing number of others see great potential in integrating public and private mobility options to better serve the public. Underlying the trend of public and private mobility integration lies two emerging concepts: Mobility-on-Demand (MOD) and Mobility as a Service (MaaS). Although MOD and MaaS differ in some aspects, both concepts focus on promoting multimodal integration, especially the integration between on-demand shared mobility and conventional public transit (Shaheen & Cohen, 2020; Smith and Hensher, 2020). The U.S. federal agencies has used the term MOD to refer to some recent initiatives that promote multimodal integration facilitated by public-private partnerships.[1]

As many researchers have argued, the development of MOD initiatives requires careful consideration of the user perspective (Calderón & Miller, 2020; Giesecke, Surakka, & Hakonen, 2016; Lyons, Hammond, & Mackay, 2019). The success of these initiatives largely depends on the willingness and capacity of travelers to use the services offered to them. Their preferences for mobility options can differ significantly, which requires nuanced policy decisions to implement MOD and to fulfill diverse travel needs. Therefore, a main research objective of this study is to reveal the taste heterogeneity three MOD-related mobility options, namely, fixed-route transit, ride-hailing services, and on-demand-based transit services (to be termed as MOD transit). Specifically, MOD transit is a concept that integrates two types of services: the large-volume mass transit (trains and buses) serves high-demand corridors, and the on-demand ride-hailing services (i.e., microtransit) serve the low-density areas and fill the first- and last-mile service gaps (Mahéo, Kilby, & Hentenryck, 2019; Stiglic, Agatz, Savelsbergh, & Gradisar, 2018; Yan, Zhao, Han, Hentenryck, & Dillahunt, 2020). We focus these mobility options for the potential substitution effects among them: as transportation agencies consider to test MOD transit pilots, it is important to know the user perceptions of MOD transit compared to those of the incumbent: fixed-route transit and ridehailing services.

Specifically, we will apply latent class cluster analysis (LCCA) to identify the latent preferences of various population segments for three shared-use mobility options: fixed-route transit, ride-hailing services, and MOD transit. LCCA allows a simultaneous analysis of the adoption and usage of the three mobility services, which can shed light on their internal relationships and provide guidance on the potential market share change once MOD transit services are introduced to the market. Moreover, by profiling latent population segments (e.g., sociodemographic characteristics, technology usage), our analysis can provide valuable insights on addressing segment-specific adoption barriers and inform targeted policy strategies to promote the adoption of MOD transit among various population groups.

This research draws on originally collected survey data from two low-income municipalities in the U.S., Detroit and Ypsilanti, Michigan. Public agencies in both localities have experimented with small-scale pilots of on-demand ride services, signaling a trend toward embracing MOD. Notably, we have employed various survey recruitment approaches such as mailing out postcards and conducting in-person recruitment at various sites to increase participation from the hard-to-reach populations, resulting in a sizeable number of disadvantaged population groups in the survey sample. The obtained survey data provide us a unique and rare opportunity to explore the implications of MOD transit services (and MOD initiatives in general) from a social equity perspective. Since the advent of MaaS and MOD concepts, many have raised equity concerns. They worry that the heavy reliance on technology makes MOD be inaccessible to less technology-savvy population groups such as low-income and older adults (Shaheen, Bell, Cohen, & Yelchuru, 2017). In addition, many current transit riders prefer cash payment, but paying for on-demand shared mobility has usually required users to have a bank account or credit card. To our knowledge, our study is one of the first empirical studies that shed light on these issues.

---

[1] According to the U.S. Department of Transportation, MOD refers to an integrated multimodal transportation network with a range of public and private travel options that are provided to travelers as an "on-demand" basis (Sheehan & Torng).

The rest of the paper is organized as follows. We begin with a review of the literature on MOD from the user perspective, focusing on the heterogeneity of MOD preferences. In Section 3, we introduce the study area and data source. We describe the method of latent class cluster analysis (LCCA) in Section 4. We present model results in Section 5 by identifying the latent segments and developing segment profiles. In Section 6, we discuss the policy implications for the MOD system. We conclude the paper with a summary of the research findings in Section 7.

## 2. LITERATURE REVIEW

MOD and MaaS are nascent ideas with much innovation and development to be experimented by transportation entities around the world. Understanding travel behavior and consumer preferences has thus attracted great research interest lately. Largely based on survey research, scholars have examined who are more likely to use MOD, how do people make tradeoffs between service attributes (e.g., the cost versus convenience), what are the attitudinal factors that impact consumer choices, and what strategies may promote adoption and use of MOD (C. Q. Ho, Mulley, & Hensher, 2020; Morsche, La Paix Puello, & Geurs, 2019; Yan, Levine, & Zhao, 2019).

Overall, the user profile of MOD adopters tends to be young, educated, tech-savvy individuals (Alonso-González, Hoogendoorn-Lanser, van Oort, Cats, & Hoogendoorn, 2020; Frei, Hyland, & Mahmassani, 2017; Vij et al., 2020), who are similar to those early adopters of ride-hailing and car-sharing services (Alemi, Circella, Mokhtarian, & Handy, 2018; Winter, Cats, Martens, & van Arem, 2020). Similar user profiles might indicate potential competitive and/or complementary relationships among these mobility services.

Despite the homogeneity of early shared mobility adopters, the MOD system can be more flexible and inclusive given its combinations of travel modes. By integrating various travel options, MOD serves a wider range of the population compared to a single travel mode. This means that MOD users are likely rather heterogenous. For example, Vij et al. (2020) identified five population segments with varying MOD [MaaS] purchase probabilities. Similar to other studies, they found young, male, college-educated, and employed individuals are most likely to adopt MOD. People from this segment have more travel needs and higher travel budgets compared to others. Meanwhile, they identified another group of middle-aged females with college degrees and high incomes, who also express a medium-to-high interest in adopting MOD. Alonso-González et al. (2020) identified five MOD preference segments. Likewise, the most likely MOD adopters tend to be young, highly educated individuals with high incomes. They also identified a group of multimodal public transport supporters as potential MOD users, who are likely to reside in urbanized areas. Ho, Hensher, Mulley, and Wong (2018) found infrequent car users are most likely to adopt MOD. Polydoropoulou, Tsouros, Pagoni, and Tsirimpa (2020) explored heterogeneity from a perspective of willingness-to-pay.

Previous studies have revealed some preference heterogeneity among MOD adopters, which provide useful insights on the acceptance of MOD initiatives across population groups. However, people from low-income communities requires more specific research attention (Jain, Ronald, Thompson, & Winter, 2017; Wong, Szeto, Yang, & Wong, 2018). Disadvantaged populations might rely more on shared-use mobility options (e.g., transit) because they often lack access to personal cars. However, low-income communities often have poor accessibility through traditional transit services (Ermagun & Tilahun, 2020). Thus, policymakers should carefully design MOD transit systems to balance the allocation of transportation resources. Moreover, people from low-income communities have diverse travel needs and preferences (Cheng, Zhao, & Li, 2019). To the authors' best knowledge, few studies have discussed taste heterogeneities among potential MOD users from low-income communities. This paper aims to fill the literature gap.

Market segmentation is a commonly used method in the transportation field to study preference heterogeneity. For instance, it has been applied to identify homogenous population groups regarding trip-making frequency (Hensher, 1976) and to analyze various inclinations in adopting autonomous vehicles (Kim, Circella, & Mokhtarian, 2019). Identifying market segments could reveal distinct behaviors and/or preferences in the population, which helps generate better understandings of the behavior and/or preference that enables proper policy suggestions towards the targeted population. Among various segmentation techniques, K-means clustering is one of the most frequently used methods (Li, Wang, Yang, & Ragland, 2013; Mokhtarian, Ory, & Cao, 2009), whereas LCCA has gained increasing prestige in recent years given its nice properties (Molin, Mokhtarian, & Kroesen, 2016; Nayum, Klöckner, & Mehmetoglu, 2016; Araghi, Kroesen, & van Wee, 2017). For example, LCCA classifies individuals with an emphasis on specific research interests while taking other related traits into considerations as well. Compared to deterministic clustering methods, the probabilistic assigning mechanism of LCCA could increase the within-segment homogeneity and between-segment heterogeneity. As such, we will apply LCCA to identify market segments.

## 3. STUDY AREA AND DATA PROCESSING

This study uses survey data from residents who live in low-income neighborhoods, focusing on their shared mobility preferences and usage. The survey was conducted in the City of Detroit and Ypsilanti Area (including the City of Ypsilanti and Ypsilanti Township), Michigan from July to November 2018. As described in a previous paper (Yan et al., 2020), both areas contain a significant proportion of populations living under poverty. By sampling from two areas, we are able to include low-income populations with various backgrounds. These low-income populations are usually underrepresented in survey data since they are hard to reach using mail or online surveys. Therefore, in addition to mail and online surveys, we conduct in-person on-site recruitment with a high cash incentive ($10 per complete response) to attract participants who may not participate in online or mailed surveys. After data cleaning and processing, we retain 825 observations for further analysis.

Table 1 presents a description of the variables used in this paper. As introduced earlier, to identify the taste heterogeneity among MOD-related mobility options, we focus on respondents' preference for three shared-use mobility services, namely, fixed-route transit, ride-hailing services, and MOD transit. Preferences for the first two mobility services are measured by the usage frequency in the last week, whereas the preference for MOD transit is measured by respondents' usage preference since the service has not yet been implemented. Here we note that variable *transit accessibility* is the first component extracted from the principal component analysis (PCA) of three transit-related variables, including the number of jobs reachable by transit in 45 min from a respondent' home, the distance of a respondent's home to the nearest bus stop, and the peak hour bus frequency at a respondent's home address. The first component captured 63.3% variance in the original variables.

**Table 1. Variables examined in this study**

| Variable | Description |
|---|---|
| Fixed-route transit usage | Indicate the usage of the fixed-route bus in the last week.<br>1 = Never used.<br>2 = Used once.<br>3 = Used twice.<br>4 = Used three or four times.<br>5 = Used five or more times. |
| Ride-hailing usage | Indicate the usage of ride-hailing services in the last week.<br>1 = Never used.<br>2 = Used once.<br>3 = Used twice.<br>4 = Used three or four times.<br>5 = Used five or more times." |
| MOD transit preference | Indicate the preference of using MOD transit over fixed-route transit.<br>1 = Strongly prefer the fixed-route over MOD.<br>2 = Sort of prefer the fixed-route over MOD.<br>3 = Not sure.<br>4 = Sort of prefer MOD over the fixed-route.<br>5 = Strongly prefer MOD over the fixed route. |
| Gender | Indicate whether the person is a male or female. |
| Race | Indicate whether the person is Black or non-Black. |
| Education | Indicate whether the person has a college degree. |
| Age | Indicate which age category the person belongs to.<br>1 = Under 25 years old.<br>2 = 25-29 years old.<br>3 = 30-39 years old.<br>4 = 40-49 years old.<br>5 = 50-59 years old.<br>6 = 60-69 years old.<br>7 = 70 years old or over. |
| Disability | Indicate whether the person has a disability. |
| Household income | Indicate which household income category the person belongs to.<br>1 = Less than $25,000<br>2 = $25,000-$49,999<br>3 = $50,000-$74,999<br>4 = $75,000-$99,999<br>5 = $100,000-$124,999<br>6 = $125,000-$149,999<br>7 = $150,000 or more |
| Vehicle ownership | Indicate whether the person owns a vehicle. |
| Cell phone ownership | Indicate whether the person has a cell phone. |
| Smartphone ownership | Indicate whether the person has a smartphone. |
| Mobile data plan | Indicate whether the person has a mobile data plan. |
| Internet access | Indicate whether the person has Internet access. |
| Transit accessibility | Indicate the person's the transit accessibility. |
| Residential location | Indicate whether the person lives in Detroit or Ypsilanti area. |

## 4. METHODOLOGY: LATENT CLASS CLUSTER ANALYSIS

This section provides a detailed description of the latent class cluster analysis (LCCA) applied here. LCCA is a probabilistic-based clustering technique that segments population groups with similar preferences and characterizes the socioeconomic profiles of these population groups. Compared to other clustering techniques, the model structure of LCCA emphasizes the variables of interests (i.e., the three mobility options in this study), and allows one to analyze the associations between key variables. Moreover, the probabilistic-based clustering mechanism introduces uncertainties when assigning individuals into different segments, generating more homogeneous segments than the deterministic-based clustering techniques. Figure 1 shows the two sub-models in the LCCA model: the membership model and the measurement model.

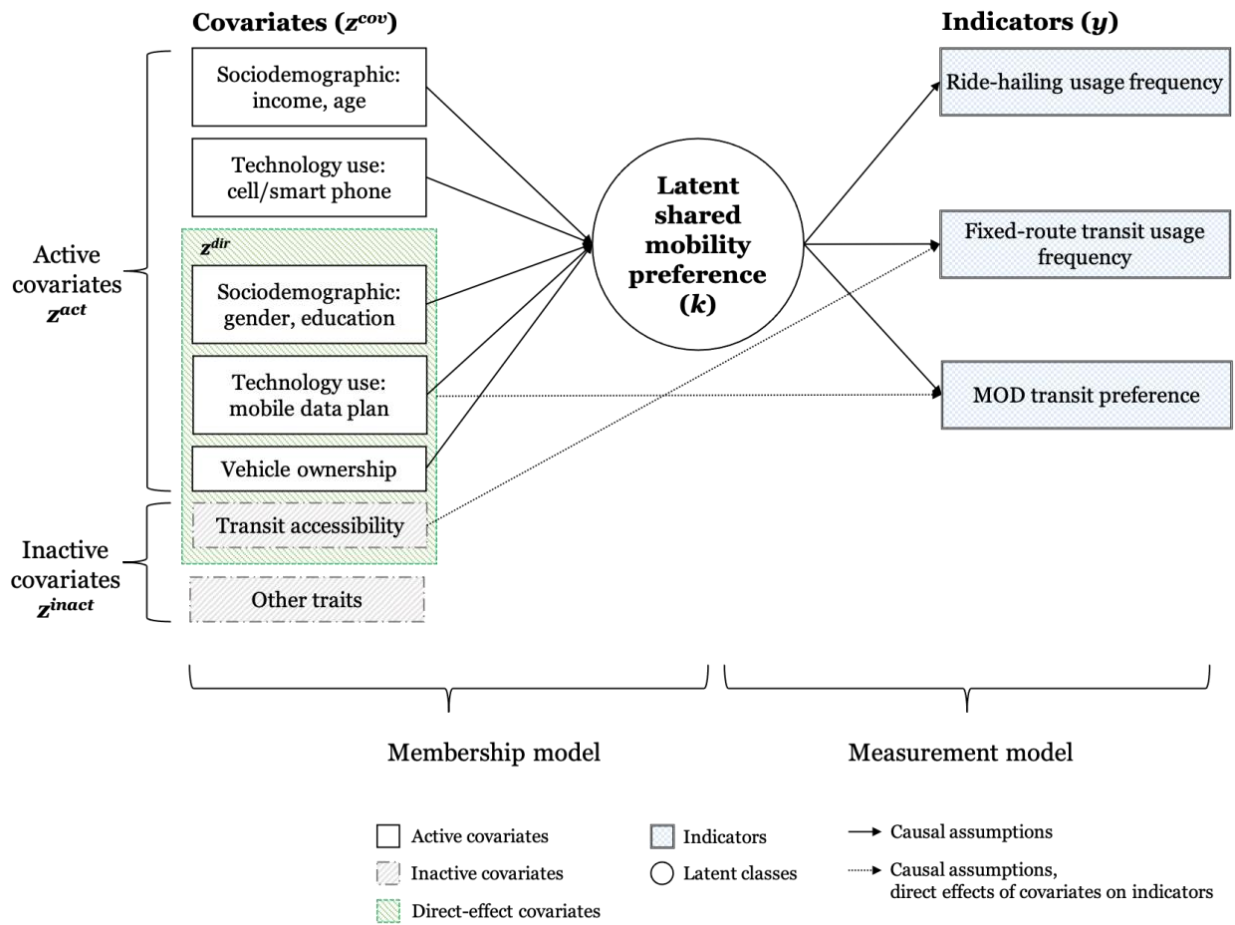

**Figure 1. Model framework of the latent class cluster analysis (LCCA)**

Specifically, the *membership model* uses active covariates ($z^{act}$) to predict the latent class membership $k$, i.e., the latent shared mobility preference segment. Active covariates include sociodemographic characteristics (e.g., income), technology use (e.g., smartphone ownership), and travel behaviors (e.g., vehicle ownership). Some of the active covariates ($z^{dir}$) such as education and mobile data plans also have direct effects on the indicator *MOD transit preference*. In other words, these covariates not only influence the classification of individuals regarding their latent shared mobility preference, but also directly influence their MOD transit preferences. Moreover, we retain covariates that are insignificant in the membership model as inactive covariates ($z^{inact}$), which do not influence the latent class structure. *Transit accessibility* is one of the inactive covariates that does not enter the membership model; however, it has direct effects on indicators *fixed-route transit usage frequency* and *MOD transit preference*. In the **measurement model**, we use the latent variable $k$ to capture the association among three observed ordinal indicators $y$: ride-hailing usage frequency, fixed-route transit usage frequency, and MOD transit preference. The three indicators reflect the current shared mobility usage of each individual as well as their preference for the new shared mobility service. Under the local independence assumption, the three indicators are mutually independent given cluster $k$.

Equations 1 to 3 are the mathematical representation of the LCCA model described above. Here, we follow the notation in Vermunt and Magidson (2016). Eq.1 represents the probability of observing the three indicators $y$ for individual $i$ given a set of observed covariates $z^{cov}$. The unobserved latent class $k$, which has $K$ categories, intervenes between $y$ and $z^{cov}$. Moreover, we also have covariates $z^{dir}$ directly influence $y$. Specifically, Eq. 1 consists of two parts. $P(k|z_i^{act})$ is the probability of the membership model and $P(y_i|k, z_i^{dir})$ is the probability of the measurement model. Given the local independence assumption, the probability of the measurement model could write as the probability product of the three indicators, i.e., $\prod_{t=1}^{3} P(y_{it}|k, z_i^{dir})$.

$$P(\mathbf{y}_i|\mathbf{z}_i^{cov}) = \sum_{k=1}^{K} P(k|\mathbf{z}_i^{act}) P(\mathbf{y}_i|k, \mathbf{z}_i^{dir}) = \sum_{k=1}^{K} P(k|\mathbf{z}_i^{act}) \prod_{t=1}^{3} P(y_{it}|k, \mathbf{z}_i^{dir}) \quad (1)$$

Eq. 2 defines the probability of individual $i$ belonging to latent class $k$ given a set of observed active covariates $z^{act}$, which is parameterized using the multinomial logit formula. For each latent class, we estimate an intercept $\gamma_{k0}$ and a set of parameters $\gamma_{kr}$ corresponding to the $R^{act}$ active covariates.

$$P(k|\mathbf{z}_i^{act}) = \frac{\exp\left(\gamma_{k0} + \sum_{r=1}^{R^{act}} \gamma_{kr} z_{ir}^{act}\right)}{\sum_{k'=1}^{K} \exp\left(\gamma_{k'0} + \sum_{r=1}^{R^{act}} \gamma_{k'r} z_{ir}^{act}\right)} \quad (2)$$

Eq. 3 defines the probability of individual $i$ with its $t$th indicator equal to $m$ given the latent class $m$ and covariates with direct effects $z^{dir}$. Please note that all three indicators are ordinal variables. As such, the probability is parameterized using the adjacent-category logit formula. We estimate an intercept $\beta_m^t$ for each ordinal value $m$, a parameter $\beta_k^t$ for each latent class, and a set of parameters $\beta_{mr}^t$ corresponding to the $R^{dir}$ covariates that directly influence the indicators. Here, the $y_m^{t*}$ is the score assigned to level $m$ of the $t$th indicator.

$$P(y_{it} = m|k, \mathbf{z}_i^{dir}) = \frac{\exp\left(\beta_m^t + \beta_k^t \cdot y_m^{t*} + \sum_{r=1}^{R^{dir}} \beta_r^t \cdot y_m^{t*} \cdot z_{ir}^{dir}\right)}{\sum_{m'=1}^{M_t} \exp\left(\beta_{m'}^t + \beta_k^t \cdot y_{m'}^{t*} + \sum_{r=1}^{R^{dir}} \beta_r^t \cdot y_{m'}^{t*} \cdot z_{ir}^{dir}\right)} \quad (3)$$

## 5. RESULTS

Similar to other clustering techniques, a modeler needs to select the optimal number of classes for the LCCA model. Since LCCA is a probabilistic-based clustering method, the model selection criterion is usually statistical measures such as Bayesian Information Criterion (BIC). In addition to model fit, interpretability should be considered when deciding the final LCCA model. In this paper, we test LCCA models with 1 to 10 latent segments. The model with three latent segments has the lowest BIC (6788.84), and the model outputs are readily interpretable. Tables 2 and 3 present the final LCCA results. To interpret the model results, we rely on segment-specific distributions of the indicators and covariates. Specifically, we first identify the three latent segments based on the segment-specific distributions of the three indicators in Section 5.1. In Section 5.2, we develop and discuss segment profiles based on the segment-specific distributions of the covariates. The results are shown in Table 4, alongside which we also present the descriptive statistics of the full sample as a reference. Lastly, for readers who are interested in technical details of the model structure, we provide our selection criterion about the direct effects (i.e., relaxed constraints) included in the LCCA model and analyze the direct effects of the covariates on indicators in Section 5.3.

**Table 2. Coefficients and z-values of the estimated LCCA measurement model (N=825)**

| $\beta_k^t$ | **Cluster 1** *Shared-mode enthusiast* | | **Cluster 2** *Shared-mode opponent* | | **Cluster 3** *Fixed-route transit loyalist* | |
|---|---|---|---|---|---|---|
| | Coefficient | z-value | Coefficient | z-value | Coefficient | z-value |
| **Ride-hailing usage freq.** | **3.76** | 3.95 | **-1.70** | -3.39 | **-2.06** | -4.05 |
| **Fixed-route transit usage freq.** | **0.35** | 5.02 | **-1.10** | -9.56 | **0.75** | 8.06 |
| **MOD transit preference** | **0.31** | 4.86 | -0.10 | -1.38 | **-0.20** | -2.17 |

| *Intercepts* $\beta_m^t$ | **Ride-hailing usage freq.** | | **Fixed-route transit usage freq.** | | **MOD transit preference** | |
|---|---|---|---|---|---|---|
| | Coefficient | z-value | Coefficient | z-value | Coefficient | z-value |
| 1 | **4.34** | 5.38 | 0.25 | 1.91 | **-1.87** | -8.95 |
| 2 | **4.97** | 4.00 | **-0.34** | -2.53 | **-0.34** | -2.94 |
| 3 | **1.69** | 5.52 | **0.35** | 4.16 | **0.56** | 6.67 |
| 4 | **-2.67** | -3.89 | **0.30** | 3.10 | **0.91** | 9.81 |
| 5 | **-8.34** | -5.08 | **-0.56** | -3.84 | **0.73** | 5.77 |

| *Direct Effects* $\beta_r^t$ | **Ride-hailing usage freq.** | | **Fixed-route transit usage freq.** | | **MOD transit preference** | |
|---|---|---|---|---|---|---|
| | Coefficient | z-value | Coefficient | z-value | Coefficient | z-value |
| **Gender** | | | | | | |
| Female | - | - | - | - | **-0.11** | -2.93 |
| Male | - | - | - | - | **0.11** | 2.93 |
| **Has a college degree** | | | | | | |
| Yes | - | - | - | - | **0.19** | 4.47 |
| No | - | - | - | - | **-0.19** | -4.47 |
| **Owns a vehicle** | | | | | | |
| Yes | - | - | - | - | **-0.19** | -3.15 |
| No | - | - | - | - | **0.19** | 3.15 |
| **Has a mobile data plan** | | | | | | |
| Yes | - | - | - | - | **0.19** | 3.55 |
| No | - | - | - | - | **-0.19** | -3.55 |
| **Transit accessibility** | - | - | **-0.07** | -2.83 | **-0.16** | -6.10 |

Notes: Coefficients use effect coding. The bolded coefficients are statistically significant at the 0.05 level.

**Table 3. Coefficients and z-values of the estimated LCCA membership model (N=825)**

| $\gamma_{k0}, \gamma_{kr}$ | Cluster 1 *Shared-mode enthusiast* | | Cluster 2 *Shared-mode opponent* | | Cluster 3 *Fixed-route transit loyalist* | |
|---|---|---|---|---|---|---|
| | Coefficient | z-value | Coefficient | z-value | Coefficient | z-value |
| **Intercept** | **1.68** | 4.09 | **-2.21** | -3.58 | 0.54 | 1.00 |
| **Gender** | | | | | | |
| Female | -0.02 | -0.28 | **0.36** | 4.18 | **-0.34** | -2.88 |
| Male | 0.02 | 0.28 | **-0.36** | -4.18 | **0.34** | 2.88 |
| **Has a college degree** | | | | | | |
| Yes | **0.54** | 4.73 | **0.33** | 2.60 | **-0.87** | -4.04 |
| No | **-0.54** | -4.73 | **-0.33** | -2.60 | **0.87** | 4.04 |
| **Age** | **-0.27** | -5.51 | 0.04 | 0.75 | **0.23** | 3.07 |
| **Household income** | **0.21** | 3.99 | 0.04 | 0.64 | **-0.25** | -2.67 |
| **Owns a vehicle** | | | | | | |
| Yes | 0.07 | 0.65 | **0.71** | 4.96 | **-0.78** | -6.04 |
| No | -0.07 | -0.65 | **-0.71** | -4.96 | **0.78** | 6.04 |
| **Has a mobile data plan** | | | | | | |
| Yes | **-0.25** | -2.13 | **0.70** | 3.58 | **-0.45** | -2.76 |
| No | **0.25** | 2.13 | **-0.70** | -3.58 | **0.45** | 2.76 |
| **Has a cell phone** | | | | | | |
| Yes | -0.01 | -0.05 | 0.79 | 1.67 | **-0.78** | -2.45 |
| No | 0.01 | 0.05 | -0.79 | -1.67 | **0.78** | 2.45 |
| **Has a smartphone** | | | | | | |
| Yes | **-0.28** | -2.27 | **0.56** | 2.82 | -0.28 | -1.56 |
| No | **0.28** | 2.27 | **-0.56** | -2.82 | 0.28 | 1.56 |

Notes: Coefficients use effect coding. The bolded coefficients are statistically significant at the 0.05 level.

## 5.1 Identification of Latent Classes

In this section, we identify and name the three latent segments based on the segment-specific mean of the three indicators, i.e., ride-hailing usage frequency, fixed-route transit usage frequency, and MOD transit preference. Figure 2 presents the segment-specific mean of the three indicators as well as the corresponding sample averages.

**Segment 1** is the largest segment, which comprises 50% of the respondents in the full sample. Individuals from this segment are frequent fixed-route transit riders (3.43, compared to sample average 2.95), and have the highest usage frequency of ride-hailing services (3.04). They also expressed the highest interest in using the MOD transit service (4.10). As such, we name Segment 1 as "**shared-mode enthusiast**".

**Segment 2** comprises 29% of the respondents in the full sample. Individuals from this segment have the lowest fixed-route transit usage frequency (1.41) among the three segments. They also rarely use ride-hailing services (1.26). Regarding the preference towards the MOD transit service, individuals from this segment have a relatively low willingness to use the MOD transit service (3.61) compared to the sample average (3.84). As such, we name Segment 2 as "**shared-mode opponent**".

**Segment 3** comprises 21% of the survey respondents. Individuals from this segment have the highest fixed-route transit usage frequency (3.95) among the three segments, whereas they have the lowest usage frequency of ride-hailing services (1.19). They also expressed the lowest interest in using the MOD transit service (3.55). As such, we name Segment 3 as "**fixed-route transit loyalist**".

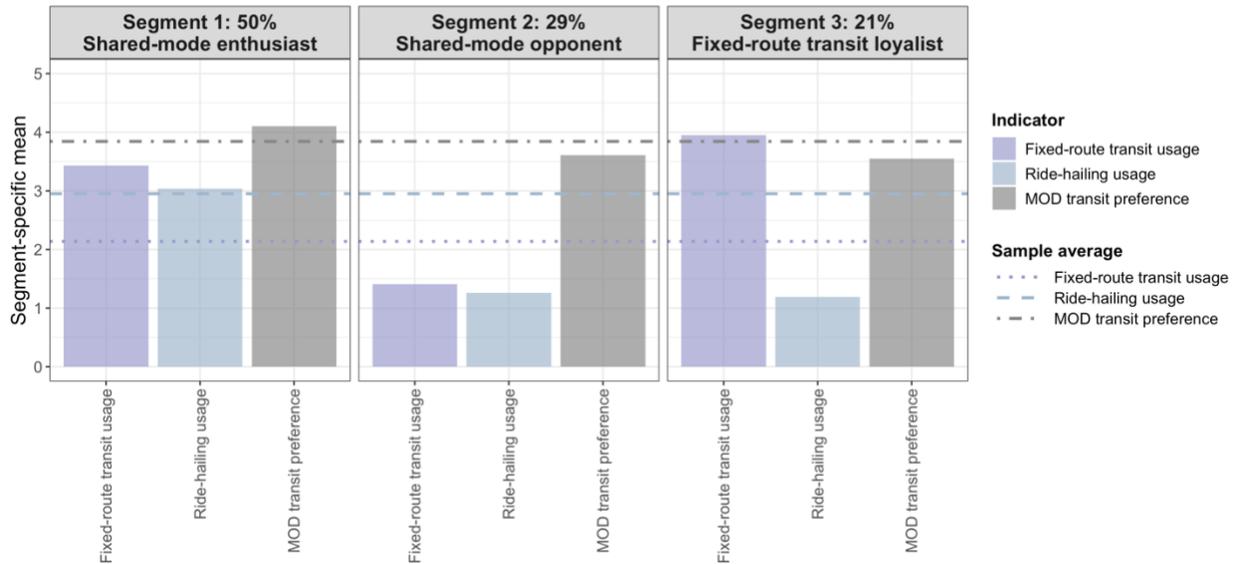

**Figure 2. Segment-specific mean for each shared mobility indicator**

### 5.2 Segment Profiles

In this section, we develop segment-specific profiles for the three latent segments. Table 4 presents the segment-specific distributions of active and inactive covariates.

The **shared-mode enthusiast** segment has a slightly larger share of males (53%) than females (47%). One-fifth of the individuals from this segment are black people, which is the lowest share among the three segments. More than half of the individuals have a college degree (64%, highest among the three segments). Moreover, people in this segment have the lowest average age, and 37% of them are younger than 30 years old. The segment also has the lowest share of people with disabilities (3%). Individuals from this segment have the highest household income. Twenty-seven percent of them have a household income above $100,000, and an equal number of individuals have a household income between $25,000 and $49,999. Moreover, a large majority of people from this segment owns vehicles (88%). Regarding technology usage, almost everyone in this segment has a cell phone (99%), and 88% of them own a smartphone. Most individuals from this segment have a mobile data plan (85%) and Internet access at home (96%). This segment has the highest share of survey respondents from Ypsilanti (37%), making this group of people have the lowest transit accessibility among the three segments (transit accessibility is generally lower in Ypsilanti than in Detroit). In general, the **shared-mode enthusiast** segment consists of college educated, technology-savvy, and young people with medium to high levels of household income.

The **shared-mode opponent** segment has the largest share of females (64%) relative to the other two segments. Roughly half of the individuals from this segment have a college degree (54%). The segment contains 29% of black people and 40% of people in the 30-39 age group. The segment contains the largest share of individuals from middle-income households (37%, $50,000 to $99,999). The segment also has the largest share of vehicle owners (95%) and individuals who have cell phones (100%), smartphones (98%), mobile data plans (97%), and Internet access (97%). Individuals from this segment live at locations with the highest transit accessibility (0.17) compared to people from the other two segments, even though they are the least frequent bus riders. More than half of the individuals in this segment live in Detroit (59%). Overall, we see that the **shared-mode opponent** segment consists primarily of middle-age, middle-income, female vehicle owners who have high technology proficiency. They have good transit access but rarely use either buses or ride-hailing services, and express low willingness towards the MOD transit service. Safety concerns may explain the travel preferences and behavior of this group of people, as it is well documented that women are often fearful of transit environments (Loukaitou-Sideris, 2014).

The **fixed-route transit loyalist** segment contains a roughly equal share of females and males (females 46%, males 54%) and the largest share of blacks (42%) relative to other segments. Only 5% of individuals from this segment have a college degree. This segment contains a large share of individuals who are aged 50 or above (39%), which contributes to the oldest average age and the largest share of disabled individuals (10%) among the three segments. Individuals from this segment also have the lowest segment-specific household income. Specifically, 62% of individuals from this segment have an income of less than $25,000. Only 33% of individuals from this segment are vehicle owners; and 18%, 21%, 22%, and 25% of the individuals do not have cell phones, smartphones, mobile data plans, or Internet access, respectively. Individuals from this segment also enjoy lower-than-average transit accessibility (-0.03 versus sample average 0.00). The majority of respondents in the segment are Detroit residents (69%, largest across the three segments). Generally, we see that people from this **fixed-route transit loyalist** segment are middle-age/older black people from low-income households who lack access to personal vehicles and the Internet. Being low income and having poor Internet access may restrict the use the ride-hailing services among this group of individuals, which also contributes to a lower preference for MOD transit.

**Table 4. Segment-specific shares/means of covariates**

| | Cluster 1<br>*Shared-mode enthusiast* | Cluster 2<br>*Shared-mode opponent* | Cluster 3<br>*Fixed-route transit supporter* | *Sample* |
|---|---|---|---|---|
| **Cluster size** | **0.50** | 0.29 | <u>0.21</u> | *1.00* |
| **Gender*** | | | | |
| Female | 0.47 | **0.64** | <u>0.46</u> | *0.52* |
| Male | 0.53 | <u>0.36</u> | **0.54** | *0.48* |
| **Black** | <u>0.20</u> | 0.29 | **0.42** | *0.27* |
| **Have a college degree*** | **0.64** | 0.54 | <u>0.05</u> | *0.49* |
| **Age*** | | | | |
| 1: Under 25 | **0.11** | <u>0.06</u> | **0.11** | *0.09* |
| 2: 25-29 | **0.26** | 0.17 | <u>0.13</u> | *0.21* |
| 3: 30-39 | 0.34 | **0.40** | <u>0.17</u> | *0.32* |
| 4: 40-49 | 0.17 | <u>0.16</u> | **0.18** | *0.17* |
| 5: 50-59 | <u>0.11</u> | <u>0.11</u> | **0.22** | *0.13* |
| 6: 60-69 | <u>0.02</u> | 0.07 | **0.16** | *0.06* |
| 7: 70 or over | <u>0.00</u> | **0.03** | 0.01 | *0.01* |
| Mean | <u>2.97</u> | 3.41 | **3.83** | *3.27* |
| **Disability** | <u>0.03</u> | 0.05 | **0.10** | *0.05* |
| **Household income*** | | | | |
| 1: Less than $25,000 | <u>0.12</u> | 0.22 | **0.62** | *0.25* |
| 2: $25,000-$49,999 | **0.27** | <u>0.24</u> | 0.26 | *0.26* |
| 3: $50,000-$74,999 | 0.19 | **0.20** | <u>0.04</u> | *0.16* |
| 4: $75,000-$99,999 | 0.16 | **0.17** | <u>0.06</u> | *0.14* |
| 5: $100,000-$124,999 | **0.14** | 0.10 | <u>0.01</u> | *0.11* |
| 6: $125,000-$149,999 | **0.08** | 0.04 | <u>0.00</u> | *0.05* |
| 7: $150,000 or more | **0.05** | 0.02 | <u>0.00</u> | *0.03* |
| Mean | **3.36** | 2.92 | <u>1.59</u> | *2.87* |
| **Own a vehicle*** | 0.88 | **0.95** | <u>0.33</u> | *0.79* |
| **Have a cell phone*** | 0.99 | **1.00** | <u>0.82</u> | *0.95* |
| **Have a smartphone*** | 0.88 | **0.98** | <u>0.79</u> | *0.89* |
| **Have a mobile data plan*** | 0.85 | **0.97** | <u>0.78</u> | *0.87* |
| **Have Internet access** | 0.96 | **0.97** | <u>0.75</u> | *0.92* |
| **Transit accessibility*** | <u>-0.08</u> | **0.17** | -0.03 | *0.00* |
| **Detroit** | <u>0.37</u> | 0.59 | **0.69** | *0.50* |

Notes: The numbers in this table represent expected values for the segment, computed with posterior class membership probabilities. Bolded numbers are the maximum segment-specific shares/means across the three segments, whereas the underlined numbers are the minimum segment-specific shares/means across the three segments. See Table 1 for variable definitions.

## 5.3 Direct effects

In this section, we provide more details of the LCCA model structure, focusing on identified direct effects of covariates on *fixed-route transit usage frequency* and *MOD transit preferences*.

As shown in Figure 1, a group of covariates including sociodemographic traits, technology usage, and vehicle ownership not only influence individual classification in the LCCA membership model, but also are directly associated with individuals' preference for the MOD transit service. Moreover, transit accessibility does not influence the classification of individuals but directly relates to individuals' fixed-route transit usage and their preference for the MOD transit service. We select these direct effects based on bivariate residuals, which is a measure of associations between observed covariates $z^{cov}$ and indicators $y$ that could not be explained by the formulated model. Specifically, we examine bivariate residuals between covariates and indicators (i.e., direct effects) and between different indicators (i.e., local dependencies). A large bivariate residual indicates a violation of the local independence assumption. By relaxing the constraints, we can expect a minimal improvement in the model fit ($L^2$ or $-2\log L$) as large as the corresponding bivariate residual (Vermunt & Magidson, 2016). As such, we relax the local independence constraints and allow associations between covariates and indicators that have bivariate residuals greater than 3.84 (i.e., the chi-square statistic at 95% significance level with 1 degree of freedom). Table 5 presents the bivariate residuals before and after relaxing local independence constraints.

**Table 5. Bivariate residuals before and after relaxing the local independence assumption**

| Covariates/Indicators | Before | | | After | | |
|---|---|---|---|---|---|---|
| | Ride-hailing usage freq. | Fixed-route transit usage freq. | MOD transit preference | Ride-hailing usage freq. | Fixed-route transit usage freq. | MOD transit preference |
| Fixed-route transit usage freq. | 0.31 | - | - | 0.36 | - | - |
| MOD transit preference | 0.69 | 0.00 | - | 1.16 | 0.08 | - |
| Gender | 0.50 | 1.36 | **11.55** | 0.34 | 1.04 | 0.00 |
| Age | 0.09 | 0.05 | 0.44 | 0.14 | 0.02 | 0.29 |
| Household income | 1.22 | 0.06 | 0.00 | 1.12 | 0.08 | 0.97 |
| Education | 0.31 | 0.73 | **16.49** | 0.60 | 0.85 | 0.00 |
| Vehicle ownership | 0.16 | 2.44 | **4.75** | 0.06 | 2.44 | 0.00 |
| Cell phone ownership | 0.01 | 1.52 | 2.02 | 0.02 | 1.70 | 0.41 |
| Smartphone ownership | 1.26 | 0.66 | 1.77 | 1.27 | 1.03 | 0.45 |
| Mobile data plan | 1.35 | 0.00 | **13.98** | 1.13 | 0.00 | 0.00 |
| Transit accessibility | 0.12 | **4.75** | **72.08** | 1.00 | 0.00 | 0.00 |

Notes: Bolded numbers are bivariate residuals greater than 3.84.

After examining the bivariate residuals, we allow direct effects of six covariates on indicators, which are direct effects of gender, education, vehicle ownership, mobile data plan, and transit accessibility on MOD transit preference as well as the direct effect of transit accessibility on the fixed-route transit usage frequency (for model coefficients, see Table 2 *Direct Effects*). We find that females are less likely to use the MOD transit service than males. This may result from females' safety concerns (Loukaitou-Sideris, 2014). Individuals who have college degrees are more likely to use the MOD transit service over fixed-route transit than people without a college degree. The result is consistent with previous findings, i.e., more educated individuals are early adopters of on-demand travel services such as ride-hailing services (Alemi et al., 2018). Moreover, people who do not have personal vehicles are more likely to choose MOD transit over a fixed-route system. Not surprisingly, lacking access to a mobile data plan is associated with a lower level of preference for MOD transit, since using on-demand travel services usually require instant communications and location sharing between users and service providers. In addition, transit accessibility has a negative relationship with the MOD transit preference. For people who can easily access to the fixed-route transit service (e.g., dense transit stops, high transit frequency), their travel demands may have been well served by the fixed-route transit system. It is natural for these individuals to resist changes (i.e., MOD initiatives) to the existing system. Unexpectedly, transit accessibility has a negative relationship with the fixed-route transit usage frequency. Usually, we expect people who live in transit-accessible neighborhoods to use transit services more frequently. We find that this counter-intuitive result is an artifact of combining data from two geographic areas: although Ypsilanti residents enjoy a lower level of transit accessibility than Detroit residents, Ypsilanti residents often use transit more as the transit services available to them are quite robust.[2]

## 6. DISCUSSION
In this section, we further analyze the segment profiles and generate policy implications for the MOD transit service.

Individuals from the **shared-mode enthusiast** segment have the highest level of interest in MOD transit among the three segments. They usually have smartphones and mobile data plans that enable them to use the MOD transit service. Given their frequent usage of both ride-hailing services and fixed-route transit, we expect that some of the shared-mode enthusiasts have already integrated the two services for their trips, while others use them frequently for separate trips. Note that although some of the individuals from this segment have relatively high income, the majority of them are from medium-to-low-income households. They are likely to choose transit over ride-hailing if transit is convenient to use. As such, MOD transit is very attractive for these individuals because of its affordability and its potential to improve transit accessibility to a large quantity and diversity of essential destinations. Overall, individuals from this segment are probably the early adopters of MOD transit services. They are likely to make use of a range of shared mobility options, whichever is available to them and convenient to use.

Individuals from the **shared-mode opponent** and **fixed-route transit loyalist** segments have lower preferences for MOD transit but for different reasons. Individuals from the **shared-mode opponent** segment are used to driving their own vehicles. They rarely use shared modes such as ride-hailing services and fixed-route transit, even though they enjoy the highest transit accessibility among the three segments on average. Pricing incentives such as ticket discounts might be a strategy to attract them to try the MOD transit services. Also, as we discussed earlier, safety concerns can prevent many females in this segment from using MOD transit. Therefore, transportation agencies should consider measures to improve the sense of safety among females.

By contrast, individuals from the **fixed-route transit loyalist** segment often do not own personal vehicles. They tend to rely on the traditional fixed-route transit to travel around. They are the so-called essential transit riders. These individuals lack interest in switching from fixed-route transit to a MOD transit system, which may be because of technological barriers or concerns for affordability. Specifically, a substantial proportion of individuals in the **fixed-route transit loyalist** segment lack access to smartphones and mobile data plans, which can prevent them from accessing MOD transit services. Transit agencies may mitigate this problem by building smart kiosks in the low-income neighborhoods to allow MOD transit reservations for people without smartphones or Internet access. Also, public agencies should evaluate discount-fare programs for MOD transit to attract the very low-income individuals.

## 7. CONCLUSION
In this study, we use the latent class cluster analysis (LCCA) to identify three latent shared mobility preference segments, namely **shared-mode enthusiast**, **shared-mode opponent**, and **fixed-route transit loyalist**. The **shared-mode enthusiast** captures highly educated young people with high or medium-level household income and high technology self-efficacy, who frequently use ride-hailing services and fixed-route buses but live in area with poor transit accessibility. They are likely early adopters of mobility-on-demand (MOD) transit services. The **shared-mode opponent** segment consists primarily of middle-age, middle-income, female vehicle owners, who rarely use shared mobility options, which is probably a result of safety concerns. Improving the safety of mobility services should thus be a priority for service providers to attract this population segment. The **fixed-route transit loyalist** segment captures middle-age and older black people from low-income households with limited access to personal vehicles. This group of individuals are often not technology savvy, which makes them feel less passionate about technology-based MOD transit services. Improving the service level of fixed-route transit as well as promoting easy and inclusive access of the MOD transit service will benefit this low-income segment. The LCCA also helps us identify variables that have direct effects on the MOD transit preference. Specifically, males, people with a college degree, carless individuals, people with mobile data plans, and people who do not have good transit access are more likely to choose MOD transit services over fixed-route transit.

---
[2] Ypsilanti and its adjacent college town, Ann Arbor, are served by the same transit operator Ann Arbor Area Transit Authority.

As public transit agencies in the U.S. and around the world embrace the trend of shifting from predominately fixed-route systems to MOD transit systems (i.e., combinations of fixed-route and on-demand-based services), the first group of beneficiaries are individuals who live in the low-transit-access area and have the habit of using ride-hailing services. Some of these people may often use ride-hailing services because public transit does not take them to where they want to go. As an MOD transit system has the potential to expand transit service areas, it could potentially serve to them as a more affordable alternative. Nevertheless, a significant proportion of transit users, labelled as **fixed-route transit loyalist** here, would require extra efforts from transit agencies to get them on board with MOD transit. Useful strategies to consider include extensive marketing and outreach activities to promote the concept of MOD transit and to educate less technology-savvy individuals how to use these services. Agencies may also facilitate easy access to MOD transit services by building smart kiosks in low-income neighborhoods. Moreover, policy measures that alleviate people's safety concerns for certain transit environments are likely to be effective to attract some female riders. Policymakers should also be aware that some people (**shared-mode opponent**) can be hard to be attracted to use shared-mobility options as they primarily travel with personal vehicles.

The paper has some limitations that warrant future research. First, the LCCA model does not include attitudinal variables such as inclinations to try new technology and to own personal cars, which could improve the model fit and bring more insights into people's travel behavior and preference. Second, this paper has focused on evaluating MOD transit preferences by measuring people's willingness to switch from a fixed-route transit system to another form publicly operated system—the MOD transit system. However, a complete MOD transport network will consist of various private and public mobility options that both complement or compete with each other. To shed light on the competition between private and public mobility services, future research should investigate people's preferences for various public transportation and privately provided shared mobility services. For instance, one could investigate whether and the degree to which people are willing to switching from privately provided ride-hailing services to publicly operated microtransit services.

## 8. ACKNOWLEDGEMENTS
This project was funded by Poverty Solutions at the University of Michigan.

## 9. AUTHORS CONTRIBUTIONS
The authors confirm contribution to the paper as follows: study conception and design: Wang, Yan, Zhao, Cao; data collection: Yan, Zhao; analysis and interpretation of results: Wang, with inputs from remaining authors; initial draft manuscript preparation: Wang, Yan, with editing from remaining authors. All authors reviewed the results and approved the final version of the manuscript.